\documentclass[letterpaper]{article}

\usepackage[T1]{fontenc}
\usepackage[utf8]{inputenc}
\usepackage{geometry}
\usepackage{setspace}

\usepackage{achemso}

\usepackage{authblk}

\usepackage{amsmath,amssymb}
\usepackage{graphicx}
\usepackage{dcolumn}
\usepackage{bm}
\usepackage{float}
\usepackage{braket}
\usepackage[normalem]{ulem}
\usepackage{multirow}
\usepackage{microtype}
\usepackage[dvipsnames]{xcolor}
\usepackage{url}    
\UseRawInputEncoding

\usepackage[
  separate-uncertainty=true,
  multi-part-units=single,
  inter-unit-product=\ensuremath{{}\cdot{}}
]{siunitx}

\DeclareSIUnit\sq{sq}
\DeclareSIUnit\T{T}
\DeclareSIUnit\dBm{dBm}
\DeclareSIUnit\mK{mK}
\DeclareSIUnit\flux{\Phi_0}

\usepackage{subfiles}
\usepackage{lineno}

\usepackage{indentfirst}
\title{Crystalline Germanium Josephson Junctions}

\author[1]{Frederik H. Knudsen}

\author[1]{Axel Leblanc}

\author[1]{Yiliu Li}

\author[1]{Patrick J. Strohbeen}

\author[1]{Jechiel van Dijk}

\author[1]{Logan Kusher}

\author[1]{Arunav Bordoloi}

\author[1]{Alisa Danilenko}

\author[2]{Xiangchao Ma}

\author[2]{Salva Salmani-Rezaie}

\author[1]{Javad Shabani\textsuperscript{*}}

\affil[1]{
Center for Quantum Information Physics,
New York University,
New York, New York, USA
}

\affil[2]{
Department of Materials Science and Engineering,
The Ohio State University,
Columbus, Ohio, USA
}

\date{
\textsuperscript{*}Corresponding author:
js10080@nyu.edu
}

\begin{document}

\maketitle

\begin{abstract}
Conventional superconducting quantum electronics rely on well-established Josephson junctions made of Al/AlO$_{x}$ where the weak link AlO$_{x}$ is amorphous and is believed to host two-level systems that limit coherence. Crystalline Josephson junctions exhibit atomically ordered interface quality but remain constrained by complex fabrication and intrinsic asymmetry of epitaxial growth. Here, we demonstrate a fully epitaxial approach based on superconductivity in gallium-doped germanium, enabling the realization of Josephson junctions entirely grown in situ by molecular beam epitaxy. These devices feature atomically sharp interfaces and crystalline weak links, resulting in strong Josephson coupling in the ultra-short regime. We observe an unconventional enhancement of the switching current under applied magnetic field, which we attribute to quasiparticle-assisted thermalization processes from the Al contacts. This platform combines structural coherence, fabrication simplicity, and scalability, offering a promising route toward low-disorder, CMOS-compatible superconducting qubits in a merged element transmon architecture.
\end{abstract}

\maketitle
\section*{Keywords}
Josephson junctions; superconductivity; germanium; crystalline semiconductor; quantum computing

\section{Introduction}

Superconducting electronics has emerged as a cornerstone of quantum technologies. At the heart of it, the Josephson junction (JJ), two superconductors separated by a weak link, brings the non-linearity required for quantum-limited amplification \cite{clerk_introduction_2010}, quantum information processing \cite{koch_charge-insensitive_2007} and ultra-sensitive detection \cite{dixit_searching_2021}. However, the amorphous junction barrier in conventional JJs is expected to host a high density of defects that will constrain device performance. This concern is most acute for merged element transmons (METs), in which most of the electromagnetic field resides within the junction itself, but it extends to conventional transmon qubits as well: now that losses in the bulk and on the surface of the superconductor have been significantly reduced, the junction interfaces and weak link have reemerged as significant loss channels \cite{bland_millisecond_2025,somoroff_millisecond_2023,rigetti_superconducting_2012}.  

To address this challenge, considerable effort has been directed towards realizing highly crystalline junctions, particularly for the tunnel barrier/weak link: vertical junctions assembled from 2D materials \cite{island_thickness_2016, vert_vdW_WMET_2025, lee_ultimately_2015, tian2021hbn}, NbN-based junctions grown by plasma-assisted molecular beam epitaxy \cite{Lachowski2025,kim_enhanced_2021}, Si-based thin-fin junctions \cite{goswami_towards_2022}. Although these approaches are highly-ordered, defining the junctions still demands multiple fabrication steps, which in turn requires aggressive cleaning or delicate handling to avoid contaminating the interfaces. 

Much of this difficulty is rooted in the asymmetry of thin film growth: if material A wets material B, favoring uniform and continuous film formation, it is often the case that material B will not wet material A. Finding a superconductor and a barrier material that are mutually compatible is therefore highly non-trivial \cite{SANDS199099}. Group-IV semiconductors such as Si and Ge offer an elegant solution: rather than pairing a superconductor with a dissimilar insulator, one can induce superconductivity directly within the semiconductor, enabling fully epitaxial super-semi heterostructures without the lattice penalty that plagues conventional insulating barriers.\cite{Shim2016}

The recent realization of superconductivity in Ga-doped Ge (Ga:Ge) realized by molecular beam epitaxy (MBE) \cite{steele_superconductivity_2025, strohbeen_superconductivity_2023} offers the possibility of growing crystalline superconducting heterostructures. Thus, Ga:Ge/Si/Ga:Ge Josephson junctions in the ultra-short regime, down to a few atomic layers, can be realized while preserving interface quality at the atomic scale. These junctions combine ease of fabrication, a crystalline weak link and a high level of tunability. They also inherit a mature CMOS-compatible Ge platform \cite{scappucci_germanium_2021, hartmann_epitaxy_2023,pitavidal2025novelqubitshybridsemiconductorsuperconductor}, on which hole spin qubits \cite{hendrickx_fast_2020, hendrickx_four-qubit_2021}, gatemons \cite{sagi_gate_2024, kiyooka_gatemon_2025} and 
cos(2$\phi$) elements \cite{valentini_parity-conserving_2024, leblanc_gate-_2025} have already been demonstrated.

Here, we report JJs fabricated from MBE-grown superconducting Ge heterostructures. \cite{steele_superconductivity_2025, strohbeen_superconductivity_2023}.
We investigate their superconducting transport and observe a strong Josephson coupling. We also find an unconventional magnetic field enhancement of the switching current, which we attribute to thermalization processes boosted by the magnetic field-induced generation of in-gap quasi-particles in the Al contacts. Finally, we discuss the potential for wafer-scale integration of superconducting qubits on this platform. Its exceptionally small footprint and low disorder make it a compelling platform for both systematic studies of two-level-system-induced loss in superconducting qubits and the realization of high-performance and CMOS-compatible quantum hardware.

\section{Results}

\subsection{Vertical crystalline Josephson junction}

\begin{figure}[!htb]
    \centering
    \includegraphics[width=1\linewidth]{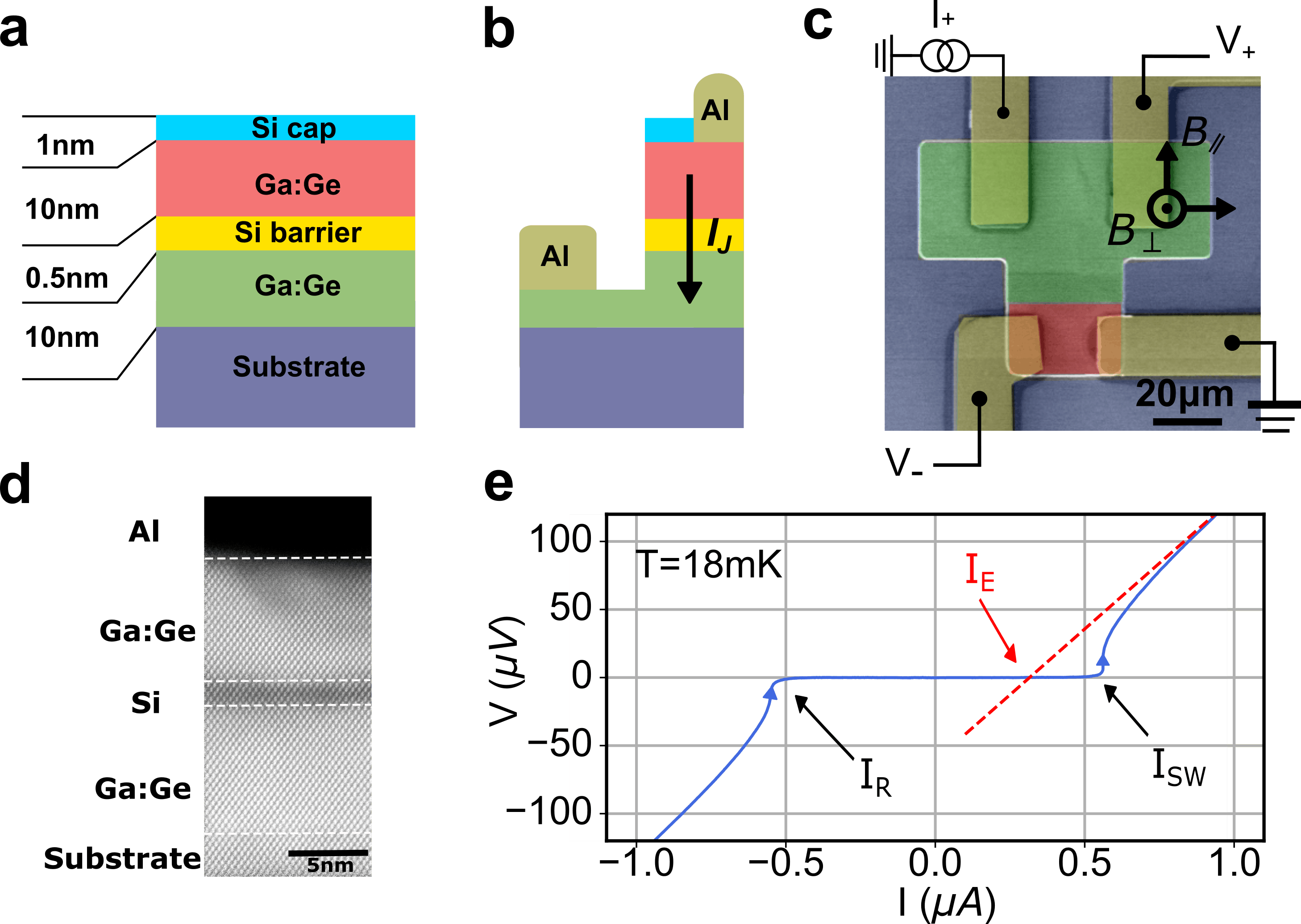}
    \caption{\textbf{A Ga:Ge vertical crystalline Josephson junction.}  \textbf{a}, The $\SI{10}{\nm}$ superconducting hyperdoped Ga:Ge films are grown on a Ge substrate and separated by a $\SI{0.5}{\nm}$ Si barrier. The heterostructure is capped by $\SI{1}{\nm}$ of Si. \textbf{b}, Superconducting Al leads contact the top and bottom Ga:Ge films, allowing transport measurements through the junction, the Josephson current $I_J$ direction is indicated by the arrow. \textbf{c}, False color Scanning Electron Microscope (SEM) image of a representative fabricated device. Four Al leads (in brown-yellow) are deposited to contact the Ga:Ge bottom layer (in green) and top layer (in red). The magnetic field orientation is as depicted here for the rest of this work. \textbf{d}, Cross-sectional STEM-HAADF image from the Al-contacted junction region, showing the coherent crystalline Ga:Ge/Si/Ga:Ge on a Ge substrate. \textbf{e}, IV characteristic of Device A at $T=\SI{18}{mK}$ measured in a four-terminal configuration by sweeping the bias current from negative to positive. Retrapping- ($I_R,$), switching- ($I_{SW}$) and excess current ($I_E$) indicated with arrows.}
    \label{fig_1}
\end{figure}

The tri-layer structure consisting of a 0.5 nm Si barrier sandwiched between two 10 nm Ga:Ge films grown on a Ge substrate using MBE to achieve superconductivity of Ge through extreme doping of Ga atoms. The Ga is substitutionally incorporated into the Ge lattice and reaching a carrier concentration of $n_h = 4.15 \times 10^{21}$cm$^{-3}$. The 0.5 nm Si barrier is coherently strained by the 4\% lattice mismatch between Ga:Ge and Si, and remains crystalline throughout the tri-layer structure. A schematic of the epitaxial material stack is shown in FIG \ref{fig_1}a. Please refer to our previous report, Ref.[\citenum{steele_superconductivity_2025}], for a comprehensive material study of the Ga:Ge/Si/Ga:Ge heterostructure. 
  
The crystalline Josephson junctions are fabricated in three lithographic steps. First, the mesa is defined using standard photolithography and etched in an inductively-coupled plasma reactive ion etching (ICP-RIE) system. To prevent the substrate from conducting due to Ga redeposition from the etched hyperdoped Ge layers, we clean the mesa-etched sample using a sequence of H$_2$O$_2$ and HCl dips. Second, we remove the top Ga:Ge layer and Si barrier on parts of the device to allow contact on both sides of the Josephson junction. This is defined and etched using the same process as the mesa. Finally, we define contacts using photolithography and deposit Al in a sputtering chamber. We deposit pairs of contacts on the top and bottom Ga:Ge layer, see FIG \ref{fig_1}b, to probe the device in a four-terminal setup. Immediately before loading the sample into the load lock of the sputtering tool, we strip the native surface oxide in buffered oxide etch 6:1, and  Cl-terminate the Ga:Ge films in HCl. See FIG \ref{fig_1}c for a Scanning Electron Microscopy (SEM) image of the final device and Supplemental Information I, for more details on device fabrication.

The atomic structure of the fabricated device was characterized by cross-sectional high-angle annular dark-field (HAADF) scanning transmission electron microscopy (STEM) in the Al-contacted junction region, as shown in FIG \ref{fig_1}d. The Ga:Ge/Si/Ga:Ge heterostructure exhibits clear lattice contrast throughout the junction, indicating that the buried interfaces remain structurally continuous after device fabrication. The corresponding elemental maps of STEM energy-dispersive X-ray spectroscopy (EDS) and normalized line profiles further confirm the expected elemental sequence across the device, as shown in FIG S1. Together, these measurements verify that the fabricated junction preserves the intended device geometry.

\begin{figure}[!htb]
    \centering
    \includegraphics[width=0.89\linewidth]{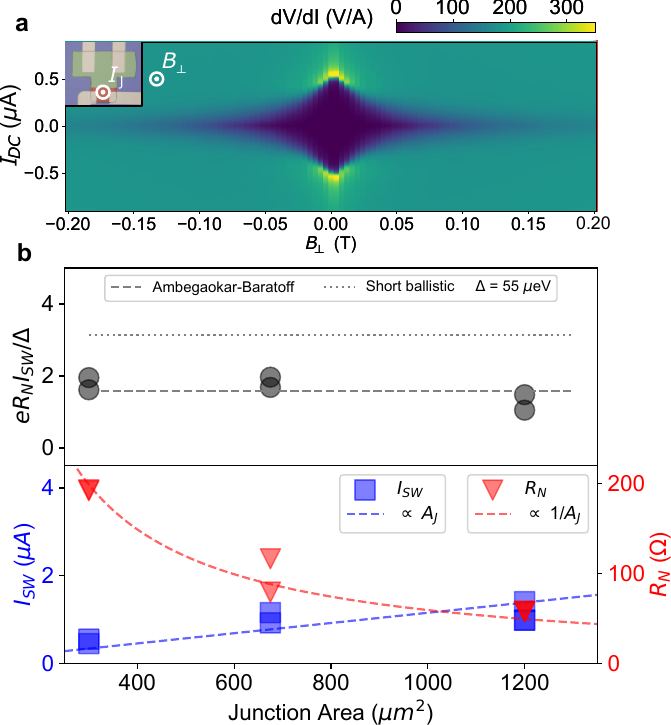}
    \caption{\textbf{Junction Characteristics.}  \textbf{a}, Differential resistance as a function of the applied magnetic field perpendicular to the substrate, B$_\perp$ (Device A). Inset: device schematic indicating the current path and field orientation. \textbf{b}, Top: $eR_NI_{SW}/\Delta$ for the six JJs with varying junction area, scattered around the Ambegaokar-Baratoff limit, and below the short ballistic limit with perfect transparency ($\tau=1$) for SNS junctions. We find that I$_{SW}$ and R$_{N}$ scales reasonably well as $A_J$ and $\frac{1}{A_J}$ respectively.}
    \label{fig_2}
\end{figure}

\subsection{Josephson junction characteristics}


For this work, we study six junctions present on the same chip. They all embed the same vertical geometry but vary in their area. To access IV characteristics, we current bias the JJ and measure the voltage drop in a four-terminal configuration so that the Al/Ga:Ge contact resistance is excluded, see FIG \ref{fig_1}c. All six devices exhibit a clear supercurrent branch; a representative characteristic is shown in FIG \ref{fig_1}e. FIG \ref{fig_2}a presents a color map of the differential resistance, dV/dI, as a function of magnetic field applied perpendicular to the substrate, B$_\perp$, for the Josephson junction. The superconducting branch exhibits its maximum critical current at zero magnetic field and narrows symmetrically with increasing perpendicular magnetic field.

The characteristics of the six devices are shown in TABLE \ref{tab:ej_ec_ratios} and we report the scaling of $I_{SW}$ and $R_{N}$ with respect to the junction area, $A_J$ in FIG \ref{fig_2}b. As expected, larger junctions can host more conduction channels and thus a larger supercurrent. In FIG \ref{fig_2}b, we also compare $eI_{SW}R_N/\Delta$ of each junction with the Ambegaokar-Baratoff limit for tunnel junctions \cite{PhysRevLett.10.486} ($eI_{SW}R_N/\Delta=\pi/2$) and the limit for short ballistic junctions ($eI_{SW}R_N/\Delta=\pi$).~\cite{kulik_josephson_1978, beenakker_josephson_1991}

\begin{table}
\centering
\begin{tabular}{ccccc}
\hline
Device & Area(\SI{}{\um\squared}) & $R_N$(\SI{}{\ohm}) & $I_{SW}$(\SI{}{\uA}) & $eR_N I_{SW}/\Delta$\\
\hline
A & 300  & 192 & 0.562 & 1.95 \\
B & 300  & 193 & 0.462 & 1.61 \\
C & 675  & 80  & 1.167 & 1.68 \\
D & 675  & 117 & 0.931 & 1.97 \\
E & 1200 & 58  & 1.414 & 1.48 \\
F & 1200 & 58  & 0.994 & 1.05 \\
\hline
\end{tabular}

\caption{\textbf{Devices summary.} All six devices present on the chip with varying junction area. The superconducting gap $\Delta$ is estimated from $T_C=$ \SI{365}{\mK} as extracted from the fit in FIG \ref{fig_4}a.}
\label{tab:ej_ec_ratios}
\end{table} 

 Although the weak link is an undoped semiconductor, we expect band bending across the ultra-short Si barrier to convert the junction from superconductor-insulator-superconductor (SIS) to superconductor-normal metal-superconductor (SNS) in character~\cite{island_thickness_2016, vert_vdW_WMET_2025}. 


\subsection{Magnetic field enhancement of the switching current and thermalization process}
\begin{figure}[!htb]
    \centering
    \includegraphics[width=0.99\linewidth]{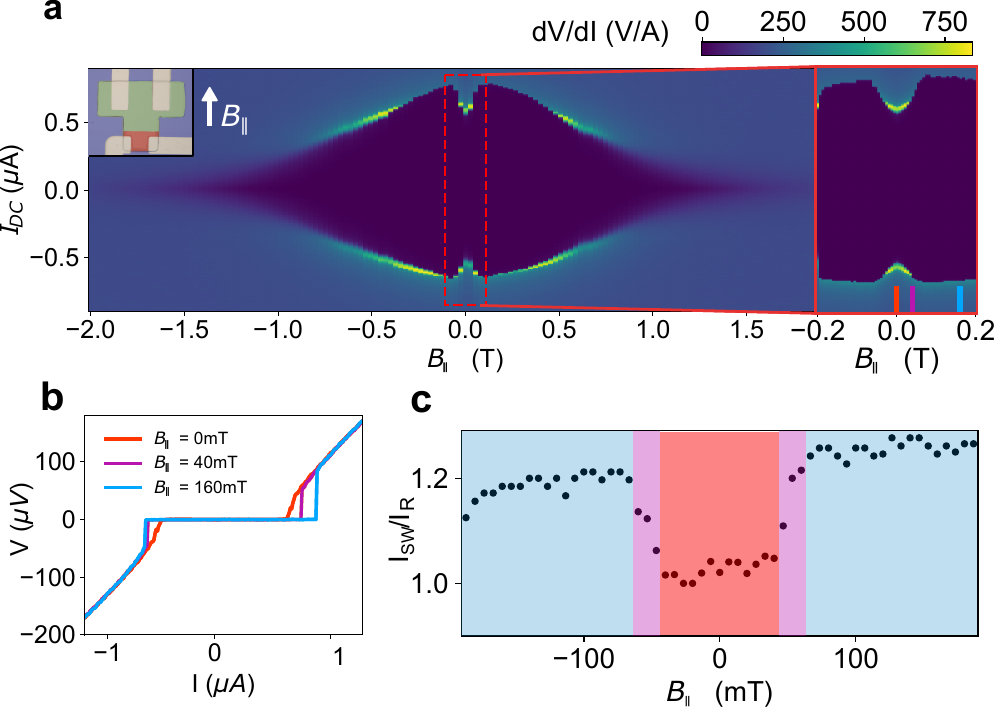}
    \caption{\textbf{Magnetic-Field-Enhanced Switching Current.} \textbf{a}, Differential resistance as a function of the in-plane magnetic field, B$_\parallel$, and the bias current, I$_{DC}$. Inset, left: device schematic indicating the field orientation. Inset, right: zoom of the region around the enhanced switching current. \textbf{b}, IV traces extracted from \textbf{a} at the field values indicated by the colored ticks. \textbf{c}, The ratio of the switching current, I$_{SW}$, to the retrapping current, I$_R$ --- a measure of IV hysteresis --- plotted against B$_\parallel$. Shaded regions mark the non-hysteretic (red), transition (purple), and hysteretic (blue) field ranges. All panels from Device A.}
    \label{fig_3}
\end{figure}

FIG \ref{fig_3}a shows the magnetic-field dependence of the junction's dV/dI characteristic, where the switching current $I_{SW}$ is enhanced around $B^*_\parallel = \SI{80}{\mT}$. FIG \ref{fig_3}b presents IV traces at $B_\parallel = 0$, \SI{40}{\mT}, and \SI{160}{\mT}, illustrating both the enhancement of $I_{SW}$ at finite field and the emergence of hysteresis. FIG \ref{fig_3}c highlights the modulation of $I_{SW}/I_{R}$, where $I_R$ is the retrapping current as a measure of the hysteresis. The enhancement of the switching current and the emergence of hysteresis are observed across all six devices on the chip.

Magnetic-field-induced enhancement of $I_{SW}$ is unusual because the critical current of a Josephson junction typically decreases with magnetic field perpendicular to the current, as shown in FIG \ref{fig_2}a. We attribute the $I_{SW}$-enhancement to enhanced thermalization of the junction by quasiparticles (QPs) assisted thermalization \cite{murani_long-_2020, wu_magnetic_2024}. The magnetic field generates in-gap QPs in the Al leads, opening a thermalization channel that complements the electron-phonon coupling to the substrate. In our geometry, this effect becomes particularly significant for the top Ga:Ge layer, which is not in contact with the substrate and relies on the Al leads as thermalization path. Consequently, we anticipate that $I_{SW}$ will reach its maximum value around the Al critical field. This maximum occurs between 65 and \SI{200}{\mT} for $B_\parallel$, which matches the Al in-plane critical field measured independently on the same chip (Supplementary Information V). However, $I_{SW}$ does not exhibit any non-monotonic behavior around zero when sweeping $B_\perp$. This could be attributed to the Ga:Ge critical field causing the vanishing of the supercurrent at lower fields than the expected enhancement due to QP generation in the Al superconducting gap.

\begin{figure}[!htb]
    \centering
    \includegraphics[width=0.8\linewidth]{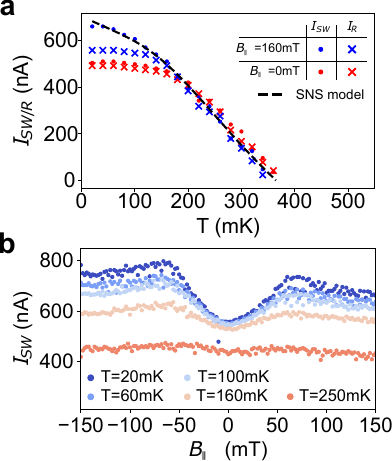}
    \caption{\textbf{Temperature Dependence.} \textbf{a}, Bath-temperature dependence of the switching and retrapping currents, $I_{SW}$ and $I_{R}$. At zero field, both $I_{SW}$ and $I_{R}$ saturate below \SI{150}{\mK}, whereas at an in-plane field $B_{\parallel} = \SI{160}{\mT}$ $I_{SW}$ continue to rise down to the base temperature while $I_{R}$ saturates, signaling the emergence of hysteresis. The dashed black line is a fit of $I_{SW}$ at $B_{\parallel} = \SI{160}{\mT}$ to a disordered short ballistic SNS model with transparency $\tau = 0.8$. \textbf{b}, Switching current $I_{SW}$ as a function of $B_{\parallel}$ at several bath temperatures. All panels show results from Device A.}
    \label{fig_4}
\end{figure}

There are several alternative mechanisms that could account for the magnetic-field-induced enhancement of $I_{SW}$, as observed in FIG \ref{fig_3}, such as magnetic impurities \cite{rogachev_magnetic-field_2006} or vortex-mediated QP trapping \cite{sato_quasiparticle_2022}. Since our system lacks magnetic impurities, we can eliminate the first hypothesis. The role of QPs can also be revealed by switching current enhancement at finite field, whether they are trapped by vortices penetrating the superconductors or whether they assist in cooling the device through the leads. The vortices picture is described by Sato et al \cite{sato_quasiparticle_2022} and supported by $I_{SW}$ hysteresis when sweeping the field up and down. However, we do not observe such magnetic field hysteresis in our device, as shown in Supplementary Information IV. We therefore exclude vortices as the source of $I_{SW}$-enhancement. Taken together, these observations make the magnetic impurities and vortex-mediated QP trapping unlikely explanations, and are consistent with QP-assisted thermalization through the Al leads as the origin of the $I_{SW}$ enhancement.

To further investigate the QP thermalization process, we measure the bath-temperature dependence of the transport. In FIG \ref{fig_4}a we report the switching $I_{SW}$ and retrapping $I_R$ currents as a function of the bath temperature for two $B_\parallel$ values, 0 and \SI{160}{\mT}. At $B_\parallel$=\SI{0}{\mT}, the switching current saturates around \SI{150}{\mK}. However, when $B_\parallel$ = \SI{160}{\mT}, saturation sets in at a significantly lower bath temperature and we notice the appearance of hysteresis in the IVs at $T<\SI{150}{\mK}$. We attribute this to a reduction in the effective electronic temperature reached when $B_\parallel$ = \SI{160}{\mT}. This picture aligns with our previous argument on the opening of a QP cooling channel at a finite field. Furthermore, a disordered short ballistic SNS JJ model \cite{beenakker_universal_1991} successfully reproduces the data at $B_\parallel$=\SI{160}{\mT} (black dashed   line) with transparency, $\tau$=0.8. The details of the model and fit are provided in Supplementary Information VI. In FIG \ref{fig_4}b, we report the $B_\parallel$ modulation of $I_{SW}$ at various temperatures. The reduction in $I_{SW}$ around $B_\parallel=\SI{0}{\mT}$ survives up to \SI{160}{\mK} and vanishes at \SI{250}{\mK}. It suggests that, above this temperature, the QP cooling process is no longer significant with respect to the cooling through electron-phonon interactions.

\section{Conclusion}
We have demonstrated crystalline Josephson junctions based on superconducting Ga-doped Ge, grown entirely in situ by molecular beam epitaxy. The resulting atomically sharp, crystalline interfaces and ultra-short weak links enable strong Josephson coupling, evidenced by low temperature transport characterization. We observe an unconventional magnetic-field-induced enhancement of the switching current that we interpret as quasiparticle-assisted thermalization from the Al contacts. These results establish doped Ge as a viable and scalable alternative to conventional amorphous-barrier junction technologies, addressing key limitations related to disorder and defect-induced decoherence.

This platform is particularly well suited to merged-element transmon (MET) architectures. Unlike the amorphous $\mathrm{AlO_x}$ barriers of standard Al-based transmons, which host large densities of two-level systems (TLS) that ultimately limit coherence, the crystalline weak link is expected to host very few TLSs. This is especially relevant for METs, where the electromagnetic energy is stored in the junction itself rather than in a large shunt capacitor, so that the enhanced electric field density inside the barrier makes TLSs in the weak link and at its interfaces even more critical. The vertical geometry and the ability to tune the weak link thickness with monolayer precision during MBE growth provide direct control over the relevant circuit parameters to reach the charge-noise-insensitive regime $E_J/E_C \approx 50$  (see Supplementary Information IX).\cite{koch_charge-insensitive_2007} Such a small physical-qubit footprint, combined with compatibility with established semiconductor processing and a fully epitaxial growth scheme, provides a route toward densely integrated, low-disorder superconducting circuits.

\bibliography{biblio}

\section*{Acknowledgments}
We acknowledge support from National Science Foundation Clemson (2137776) and Air Force Office of Scientific Research Multidisciplinary University Research Initiative (FA9550-25-1-0289). Electron microscopy was performed at the Center for Electron Microscopy and Analysis (CEMAS) at The Ohio State University. S.S.-R. acknowledges D. Huber for assisting with the TEM sample preparation. This work was performed in part at the Nanofabrication Facility at the Advanced Science Research Center at The Graduate Center of the City University of New York.

\section*{Author contribution}
Manuscript preparation: A.L. and F.H.K., with assistance from Y.L., J.S. and A.B. Device fabrication: F.H.K., with assistance from L.K., Y.L. and A.D. Sample growth: P.J.S., with assistance from J.v.D. Transport measurements: A.L. and F.H.K. Electron microscopy (SEM, STEM, STEM-EDS): X.M. and S.S.-R.

\section*{Competing interests}
The authors declare no competing interests.

\section*{Data availability}
All data used to produce the figures in this paper and to support our analysis and conclusions are available upon request.

\end{document}



\title{Supplementary Information}

\maketitle

\section{Fabrication}

The mesa was defined using AZ 5214-E photo resist spin coated at 4000 RPM for 60 seconds and exposed in a Karl Suss MJB3 mask aligner. The mesa is etched in an inductively-coupled plasma reactive ion etcher (ICP-RIE, Oxford PlasmaPro System 100 Cobra) using an etch chemistry consisting of BCl$_3$ (5 sccm), Cl$_2$ (0.5 sccm) and Ar (5 sccms). The ICP and table RF power is set to 500W and 10W respectively. The etch depth for the mesa is 90 nm. To prevent the substrate from conducting due to redeposited Ga from the etched hyperdoped Ge layers, we clean the mesa etched sample. This is done in four cycles of “digital etching”: 10 seconds in H$_2$O$_2$ (30\%), DI Water dip, 10 sec in HCl (37\%) and DI water dip. The photo resist is then stripped in Remover PG  at \SI{50}{\degreeCelsius} for 30 minutes with gentle ultra sonication of 10 seconds every 10 minutes. 

Second, we open the bottom layer of Ga:Ge on parts of the device to allow contact on both sides of the Josephson junction. This is defined and etched using the same process as the mesa, only with the etch time reduced. The etch depth has to be very accurate: A few Angstrom too shallow, and the top film is shorted across the device, or a few angstrom too deep, and the bottom Ga:Ge is damaged. To mitigate for variations in the etch chamber between batches, we use a same day calibration protocol, where the mesa is etched and measured the same day as the second etch, to achieve consistent precision down to a few Angstrom. 

Finally, we define contacts using photolithography and deposit Al in a sputtering chamber (AJA Orion 8). Immediately before loading the sample into the load lock of the sputtering tool, we strip the oxidized 1 nm Si cap in a 10 seconds dip in buffered oxide etch 6:1, followed by a 10 seconds dip in HCl to Cl-terminate the Ga:Ge films and prevent oxidation. The transfer time is kept below 4 minutes. Lift-off is done in acetone at \SI{50}{\degreeCelsius} for 1 hour, followed by gentle sonication. We note that there is no spacer between the top contact and the bottom Ga:Ge, which could result in a short, although this is deemed highly unlikely to occur, since we have an anisotropic, dry-etched mesa and a 10 nm Ga:Ge that is partially oxidized. Furthermore, we confirm during I-V characterization that there is no such short.

\begin{figure}[h]
    \centering
    \includegraphics[width=0.7\linewidth]{Fig/Fig (Supp)/EDS_combined_fig.png}
    \caption{\textbf{STEM-EDS}  \textbf{a.} Individual EDS intensity maps of the Ge (Blue), Ga (Red), Si (Green) and Al (Yellow), as well as the superimposed map of all four elements. \textbf{b.} The normalized EDS line profiles from the maps in panel.} 
    \label{EDS}
\end{figure}

\section{STEM characterization}
Cross-sectional STEM lamellae were prepared from JJs using focused ion beam (FIB) milling. Initial thinning was performed at 30 kV, followed by final low-voltage polishing at 5 kV to reduce FIB-induced surface damage and amorphization. STEM was carried out on a Thermo Fisher Scientific Themis Z microscope operated at 200 kV with a probe semi-convergence angle of 20 mrad. High-angle annular dark-field STEM images were acquired using a detector collection range of 64 to 200 mrad. To improve the signal-to-noise ratio while minimizing scan distortions, 20 fast-scan images were acquired with 2048 × 2048 pixels and a dwell time of 200 ns, then aligned and averaged. Energy Dispersive X-ray Spectroscopy (EDS) spectrum imaging was performed beneath the contact region using a Super-X detector (FIG \ref{EDS}).

Cross-sectional STEM was performed on FIB lamellae prepared from fabricated junction devices to verify the layer sequence after processing. In the Al-contacted junction region, the nominal Al/Ga:Ge/Si/Ga:Ge/Ge vertical stack is observed. The Ga:Ge layers retain clear lattice contrast, and the buried junction region remains structurally continuous. STEM and EDS maps and normalized line profiles confirm the expected elemental sequence across the fabricated device structure. These measurements verify that the processed junction retains the intended device geometry, while detailed structural characterization of the as-grown Ga:Ge heterostructure has been reported previously \cite{steele_superconductivity_2025}.

\section{IV Characterization}

\begin{figure}[h]
    \centering
    \includegraphics[width=0.9\linewidth]{Fig/Fig (Supp)/all_IVs.pdf}
    \caption{\textbf{Comparison of IV curves}  \textbf{a.} The IV curves of the six JJs (A-F) measured with current bias in a 4-probe configuration with no applied field. \textbf{b.} The IV curves for the same six devices, now with applied in-plane magnetic field, $B_\parallel = \SI{100}{}mT$. \textbf{c.} The product $eI_CR_N/\Delta$ of all 6 devices as a function of the devices area. The right axis displays the corresponding transparency in a short ballistic SNS junction framework.}.
    \label{AllD}
\end{figure}

In this work, we have studied six JJs on the same chip, all of which shows a supercurrent branch, see FIG \ref{AllD}a. With an applied in-plane magnetic field we see enhanced $I_C$ and hysteretic behavior, FIG \ref{AllD}b. We note that Device B and Device F does not show significant hysteretic behavior and overall reduced $I_C$ compared to the other JJs with the same area. FIG \ref{AllD}a, displays the computed $eI_CR_N/\Delta$ product of each junction as a function of the areas whit $R_N$ measured at large bias, $\Delta$ estimated from the junction critical temperature reported in FIG.4a. The scaling to transparency $\tau$ is done in the SNS short ballistic framework following $eI_CR_N/\Delta = \pi / (1+\sqrt{1-\tau})$.

\section{Magnetic Field Hysteresis and Vortex Dynamics}
It has been suggested that enhancement of the supercurrent at finite applied magnetic field could be due to QP trapping by superconducting vortices \cite{sato_quasiparticle_2022}. The QP bound states at the vortex cores have lower energy, which allows a new mode of QP relaxation that effectively reduces the electronic temperature in the process. While the enhancement of $I_{SW}$ from vortex cooling appears to be very similar to what we observe for our JJ devices, we rule out this explanation for two reasons:

\textbf{1)} The Al contacts in our device are only expected to act as a type II superconductor for B-fields applied out-of-plane. For fields larger than the critical field, $B_{c_1}^{\perp}$, the contacts have magnetic fields penetrating the film which host vortices. However, we do not see any noticeable enhancement of $I_{SW}$ in $B_\perp$. On the contrary, we see enhancement for finite $B_\parallel$ where no vortices are expected to emerge. While this argument rules out vortex induced cooling from the Al leads, we cannot exclude the possibility of the Ga:Ge films to host vortices that emerge at finite $B_\parallel$.

\textbf{2)} To address the possibility of $B_\parallel$-induced vortices in the Ga:Ge film, we sweep the B-field in different directions, see FIG \ref{HystB}. Ramping down from large fields, the vortices are expected to remain in the film before vanishing at a field significantly smaller than $B_{c_1}$. However, we see no hysteretic behavior for the switching current and we therefore rule out the possibility of vortex-induced cooling effects in any of the two superconducting films.

\begin{figure*}[]
    \centering
    \includegraphics[width=0.95\linewidth]{Fig/Fig (Supp)/B_Hyst.pdf}
    \caption{\textbf{Hysteresis-free Switching Current for Applied Magnetic Field}. The switching current, $I_{SW}$, plotted against in-plane magnetic field, $B_{\parallel}$, at different temperatures and for both magnetic field sweep directions: ramped up (blue) and ramped down (red). For all temperatures, we find that the magnetic field show no hysteretic behavior of $I_{SW}$, eliminating vortices as a dominant cooling effect through QP trapping.}
    \label{HystB}
\end{figure*}

\section{Aluminum critical field}

The critical current enhancement is expected to be maximal when the Al contact superconducting gap closes. In order to test our hypothesis, we conduct a critical field measurement of the Al deposited on the chip. In FIG. \ref{Al_Bc} we report the temperature dependence of in plane $B^C_\parallel$ and out of plane $B^C_\perp$ critical field of the Al.

\begin{figure*}[]
    \centering
    \includegraphics[width=0.6\linewidth]{Fig/Fig (Supp)/Al_Bc.pdf}
    \caption{\textbf{Aluminum in plane and out of plane critical field}.}
    \label{Al_Bc}
\end{figure*}

\section{Fitting of $I_C(T)$}

In FIG. 4a, we show the switching and retrapping current with respect to temperature at $B_\parallel$ =0 and \SI{160}{\mT}. We try to fit the switching current for both fit with a short disordered ballistic model \cite{beenakker_josephson_1991,golubov_current-phase_2004} accounting for $N$ conduction channels and an overall effective transparency $\tau$:
\begin{equation}
    I_{C}(T) = \max_{\phi}\left[ \sum_N \frac{e\Delta(T)}{2\hbar}\frac{\tau\sin(\phi)}{\sqrt{1-\tau\sin^2(\phi/2)}}\tanh{\frac{\Delta(T)}{2k_BT}\sqrt{1-\tau\sin^2(\phi/2)}}\right]
\end{equation}
where $\phi$ is the superconducting phase difference across the junction, $\Delta(T)=1.74\Delta_0\sqrt{1-T/T_C}$ is the temperature dependent superconducting gap of the leads, $T_C$ being its superconducting critical temperature. The fit cannot reproduce the data for $B_\parallel$ =\SI{0}{\mT} because of the $I_{SW}$ saturation at low temperature due to poor thermalization. However, at $B_\parallel$ =\SI{160}{\mT}, the fit successfully converges to $\tau=0.8\pm 0.07$, $N=54\pm8$ and $T_C=$\SI{365\pm 5}{\mK}.

We note here that $N=54$ conduction channels with $\tau=0.8$ would correspond to a normal resistance $R_N$=\SI{298}{\ohm} following the Landauer formula:
\begin{equation}
    R_N = \frac{1}{N}\frac{h}{2\tau e^2}
\end{equation}
However, we measure a lower normal resistance $R_N$=\SI{191}{\ohm} suggesting the presence of more conduction channels that doesn't participate in a significant way to the supercurrent transport potentially due to their lower transparency.

\begin{figure}
    \centering
    \includegraphics[width=0.8\linewidth]{Fig/Fig (Supp)/circuit.pdf}
    \caption{\textbf{Circuit Diagram of the Measurement Setup} Simultaneous measurement of DC and differential resistance in four-probe configuration with current bias: $R_{DC}=\SI{200}{k\Omega}$ and $R_{AC}=\SI{1}{M\Omega}$.}
    \label{cir}
\end{figure}

\section{Measurement Setup}
The four-terminal measurement scheme is shown in FIG \ref{cir}. Standard IV-curves are captured from the DMM, and the differential resistance, $dV/dI$, is recorded directly from the SR860 lock-in amplifier by demodulation. The SR860 source the alternating current with excitation of \SI{10}{mV} over a \SI{1}{M \ohm} bias resistor at \SI{117}{Hz}.  

The four-probe measurement scheme isolates the JJ and ensures that we do not capture the voltage difference across external features such as the Al contacts. 

\section{Comparison: Magnetic Field Dependence}
While the main text focuses on Device A for the magnetic field dependence, we compare here the magnetic field dependence between devices A, C and E, see FIG \ref{Bcomp}. The plots show a consistent pattern: for finite $B_\parallel$, $I_{SW}$ is enhanced while $I_{SW}$ monotonically decreases with applied $B_\perp$. See the main text for detailed discussion of these phenomena.

\begin{figure}
    \centering
    \includegraphics[width=0.8\linewidth]{Fig/Fig (Supp)/All_Dev_Bmaps.pdf}
    \caption{\textbf{Magnetic Field Dependence Across Devices.} Magnetic field in in-plane, $B_\parallel$ and out-of-plane, $B_\perp$ orientations for devices A, C and E. The differential resistance, $dV/dI$ is captured directly from the lock-in amplifier (see FIG \ref{cir}), except for Dev. E, $B_\perp$ which has been derived numerically from its IV-trace. }
    \label{Bcomp}
\end{figure}

\section{Junction Area for Transmon Qubit}
Standard amorphous aluminum oxide tunnel barrier commonly used in Al-based transmon qubits are known to host large densities of two-level systems (TLS), which introduce decoherence channels that ultimately may limit qubit performances. In contrast, this single crystalline JJ is expected to host few TLSs, making it a promising platform to clarify their impact on coherence performances. Furthermore, the vertical geometry and the ability to precisely tune the weak link thickness during the MBE growth opens the possibility of engineering METs, in which the electromagnetic field energy is primarily stored in the JJ itself rather than in a large shunted capacitor, as in standard transmons. In such a design the electric field density is higher inside the junction making TLSs present in the weak link and at the interfaces even more critical. This motivates the choice of an in-situ crystalline JJ. The charge noise-insensitive regime where the Josephson energy ${E_J}$ and charging energy ${E_C}$ satisfy $\frac{E_J}{E_C}\approx50$ is realized by having a sufficiently large capacitance from the junction itself \cite{koch_charge-insensitive_2007}. As a crude approximation of $\frac{E_J}{E_C}$, we model the junction as a parallel plate capacitor with relative permittivity of Si, $\epsilon_r = 11.7$ \cite{Si1956} (Diffusion of Ga into the Si barrier has not been included in the estimate of the capacitance, although the EDS suggests that this could be the case).  

For a parallel plate barrier we have $\frac{E_J}{E_C} \propto A^2$. As such, interpolation of the JJ data predicts that the area must be reduced to \SI{2.96\pm0.15}{\micro\meter\squared} to reach $\frac{E_J}{E_C}=50$, see FIG. \ref{EJEC}. Such a small physical-qubit footprint suggests a route to integrating many qubits on a single chip.

\begin{figure}
    \centering
    \includegraphics[width=0.5\linewidth]{Fig/Fig (Supp)/EJEC.pdf}
    \caption{\textbf{Tuning $E_J/E_C$ with Junction Area} The ratio between junction energy and charging energy, $E_J/E_C$ for JJ devices (blue) with different junction areas. The fit $E_J/E_C \propto A^2$ (grey) intersects the energy ratio for the transmon regime, $E_J/E_C=50$ (red) at \SI{2.96\pm0.15}{\micro\meter\squared} (green). Inset: enlarged plot in the region around the intersection where the error on $E_J/E_C$ is propagated from the fit.}
    \label{EJEC}
\end{figure}

\bibliography{biblio}